\documentstyle[12pt,epsfig]{article}

\input axodraw.sty

\newcounter{minutes}
\def\lsim{\:\raisebox{-0.5ex}{$\stackrel{\textstyle<}{\sim}$}\:}
\def\gsim{\:\raisebox{-0.5ex}{$\stackrel{\textstyle>}{\sim}$}\:}

\newcommand{\newc}{\newcommand}
\newc{\pbi}{pb$^{-1}$}
\newc{\ie}{{\it i.e.} }
\newc{\ti}{\tilde}
\newc{\ra}{\rightarrow}
\newc{\ee}{$e^+e^-$\ }
\newc{\mm}{$\mu^+\mu^-$}
\newc{\taus}{$\tau^+\tau^-$}
\newc{\uu}{$u\bar{u}$\ }
\newc{\eeee}{$e^+e^-\ra e^+e^-$\ }
\newc{\eemm}{$e^+e^-\ra \mu^+\mu^-$\ }
\newc{\eett}{$e^+e^-\ra \tau^+\tau^-$\ }
\def\Rs{R \hspace{-0.38em}/\;}
\newc{\beq}{\begin{eqnarray}}
\newc{\eeq}{\end{eqnarray}}
\newc{\dqu}{\delta_{qu}}
\newc{\dqd}{\delta_{qd}}
\newc{\non}{\nonumber}
\newc{\noi}{\noindent}

\def\ib#1,#2,#3{       {\it ibid.\/ }{\bf #1} (19#2) #3}
\def\ap#1,#2,#3{       {\it Ann.~Phys.~(NY)\/ }{\bf #1} (19#2) #3}
\def\ijmp#1,#2,#3{     {\it Int.\ J.~Mod.\ Phys.\/ } {\bf A#1} (19#2) #3}
\def\mpl#1,#2,#3 {     {\it Mod.~Phys.~Lett.\/ } {\bf A#1} (19#2) #3}
\def\npb#1,#2,#3{       {\it Nucl.\ Phys.\/ }{\bf B#1} (19#2) #3}
\def\npps#1,#2,#3{     {\it Nucl.\ Phys.~B (Proc.~Suppl.)\/ }{\bf B#1}
                             (19#2) #3}
\def\plb#1,#2,#3{      {\it Phys.\ Lett.\/ }{\bf B#1} (19#2) #3}
\def\pr#1,#2,#3{       {\it Phys.\ Rev.\/ }{\bf #1} (19#2) #3}
\def\prd#1,#2,#3{      {\it Phys.\ Rev.\/ }{\bf D#1} (19#2) #3}
\def\prep#1,#2,#3{     {\it Phys.\ Rep.\/ }{\bf #1} (19#2) #3}
\def\prl#1,#2,#3{      {\it Phys.\ Rev.\ Lett.\/ }{\bf #1} (19#2) #3}
\def\pro#1,#2,#3{      {\it Prog.~Theor.\ Phys.\/ }{\bf #1} (19#2) #3}
\def\rmp#1,#2,#3{      {\it Rev.~Mod.~Phys.\/ }{\bf #1} (19#2) #3}
\def\sp#1,#2,#3{       {\it Sov.~Phys.~Usp.\/ }{\bf #1} (19#2) #3}
\def\zpc#1,#2,#3{      {\it Z.~Phys.\/ }{\bf C#1} (19#2) #3}
\def\appb#1,#2,#3{     {\it Acta Phys.\ Polon.\/ }{\bf B#1} (19#2) #3}

\begin{document}

\begin{flushright}
  BI-TP 97/08\\ DESY 97-044\\ WUE-ITP-97-04\\[1.7ex] {\tt
    hep-ph/9703436} \\ 
\end{flushright}

\vskip 35pt
\begin{center}
  {\Large \bf Supersymmetry with $R$-Parity Breaking: }\\[2mm] 
  {\Large \bf Contact Interactions and Resonance Formation }\\[2mm] 
  {\Large \bf in Leptonic Processes at LEP2 }

\vspace{5mm} 
{\large J. Kalinowski}$^{1,2}$, 
{\large R. R\"uckl}$^{3,\displaystyle \ast}$, 
{\large H. Spiesberger}$^{4,\displaystyle \ast}$,\\[1.1ex] 
{\large and P.M. Zerwas}$^1$\\[2ex] 
{\em $^1$ Deutsches Elektronen-Synchrotron DESY, D-22607
  Hamburg}\\[1.1ex]  
{\em $^2$ Institute of Theoretical Physics, Warsaw University,
  PL-00681 Warsaw}\\[1.1ex] 
{\em $^3$ Institut f\"ur Theoretische Physik, Universit\"at
  W\"urzburg, D-97074 W\"urzburg}\\[1.1ex] 
{\em $^4$ Fakult\"at f\"ur Physik, Universit\"at Bielefeld, D-33501
  Bielefeld}\\[2ex]


\vspace{1cm} {\bf ABSTRACT}
\end{center}

\begin{quotation}
  In supersymmetric theories with $R$-parity breaking, trilinear
  couplings of two leptons to scalar sleptons are possible.  In
  electron--positron collisions such interactions would manifest
  themselves through contact terms in Bhabha scattering, $e^+e^-
  \rightarrow e^+e^-$, and in annihilation to lepton pairs, \eemm and
  \taus.  Interpreting the high $x$, high $Q^2$ DIS HERA events as
  $charm$ squark production with squark masses of order 200 GeV, the
  formation of $tau$-sneutrinos, $e^+e^-\ra \ti{\nu}_{\tau}$, with a
  mass in the range close to the LEP2 energy or even in reach, is an
  exciting speculation which can be investigated in the coming LEP2
  runs with energies close to $\sqrt{s}=200$ GeV.
\end{quotation}

\vspace*{\fill} \footnoterule {\footnotesize
  \noindent ${}^{\displaystyle \ast}$ Supported by Bundesministerium
  f\"ur Bildung, Wissenschaft, Forschung und Technologie, Bonn,
  Germany, Contracts 05 7BI92P (9) and 05 7WZ91P (0).}

\newpage \renewcommand{\thefootnote}{\arabic{footnote}}


\section{Introduction}
The recent observation of surplus events in deep-inelastic
positron--proton scattering at HERA at high $x$ and high $Q^2$ above
{\it a priori} expectations \cite{sem} has given rise to many
speculations.  If the surplus is not a statistical fluctuation, an
attractive interpretation is offered by supersymmetry with $R$-parity
breaking \cite{chokal}. Since in particular the H1 events
cluster at a mass value of 200 GeV, resonance squark production
$e^+d\ra\ti{c},\ti{t}$ could explain the HERA
events\footnote{Neutrinoless double $\beta$ decay \cite{beta}
  restricts the $e^+d\ti{u}$ coupling so strongly that this
  interaction cannot account for the $(ej)$ final states at HERA.}
without spoiling the tremendous success of the high-precision analyses
based on  the Standard Model.

In addition to the lepton-quark-quark superfield term, the $R$-breaking
part of the superpotential may involve also the interaction of three
lepton superfields \cite{Rbroken,suprp}:
\beq
W_{\Rs}=\lambda_{ijk}L^i_LL^j_L\bar{E}^k_R +
\lambda'_{ijk}L^i_LQ^j_L\bar{D}^k_R \label{superp} 
\eeq 
Both couplings $\lambda$ and $\lambda'$ violate lepton number
($L$). Their coexistence is not excluded by the non-observation of
proton decay.  The indices $ijk$ denote the generations; 
$\lambda_{ijk}$ are non-vanishing only for $i < j$ so that
at least two different generations are coupled in the purely leptonic
vertices.  The standard notation is used in Eq.\ (\ref{superp}) for
the left-handed doublets of leptons ($L$) and quarks ($Q$), and the
right-handed singlets of charged leptons ($E$) and down-type quarks
($D$). In four-component Dirac notation, the lepton part of the Yukawa
interactions has the following form:
\beq 
{\cal L}_{\Rs}^l=\lambda_{ijk}\left[
\ti{\nu}^j_L\bar{e}^k_Re^i_L 
+(\ti{e}^k_R)^*(\bar{e}^i_L)^c\nu^j_L
+\ti{e}^i_L\bar{e}^k_R\nu^j_L
-\ti{\nu}^i_L\bar{e}^k_Re^j_L
-(\ti{e}^k_R)^*(\bar{e}^j_L)^c\nu^i_L 
-\ti{e}^j_L\bar{e}^k_R\nu^i_L
\right] + h.c.  
\eeq 
$u^i$ and $d^i$ denote the  $u$- and $d$-type quarks, $e^i$ and $\nu^i$
the charged and neutral leptons, respectively; $\bar{l}$ denotes the
spinor of the antiparticle, the superscript $(~)^c$ the charge
conjugate spinor and $(~)^*$ the complex conjugate scalar.

The interpretation of the HERA events by $R$-parity breaking SUSY
interactions involves at least one of the couplings $\lambda'$, in the
most attractive scenarios $\lambda'_{121}$ or $\lambda'_{131}$, giving
rise to {\it charm} or {\it top} squark production with masses $\sim
200$ GeV, respectively.  This invites to the speculation that some of
the couplings $\lambda$ may also be non-zero in the purely leptonic
sector and that other supersymmetric particles, sleptons, may exist in
a similar mass range.  A similar idea has been envisaged \cite{cglw}
in the charged slepton sector to account for the Aleph 4-jet events
\cite{aleph}.  In the present paper we investigate sneutrino effects
in leptonic $e^+e^-$ processes at the high energies\footnote{Other
  novel interactions which may be $l-q$ symmetric \cite{wilbar}, could
  give rise to effects in the lepton sector which are similar to the
  effects in supersymmetric theories with $R$-parity breaking.}
realized at LEP2.  They include Bhabha scattering and $\ell^+\ell^-$
pair production:
\begin{eqnarray}
e^+e^-& \ra& e^+e^- 
\label{bhabha}
\\ e^+e^-& \ra & \mu^+\mu^-,\;
\tau^+\tau^- 
\label{mutau} 
\end{eqnarray}
Neutrino pair-production, involving the exchange of charged sleptons,
can be analysed in the same way after obvious substitutions, though
experimental analyses are much more difficult.  Both processes
(\ref{bhabha}) and (\ref{mutau}) can be affected by the exchange of
sneutrinos in the $s$- and/or $t$-channel.  For sneutrinos with masses
in the order of 200 GeV, the effects can be quite significant,
depending on the size of the couplings. Even though there are strong
upper bounds on several of the $\lambda$ couplings, some of these
couplings are rather unconstrained, in particular the coupling that
violates only the $\tau$-flavor, so the effects induced by $\tau$
sneutrinos can be large. While contact interactions relevant for much
heavier sneutrinos have been discussed earlier in the literature
\cite{DH,CC,godbole}, we improve on these analyses by including the
impact of nearby resonances; they require the proper account of
sneutrino propagator and non-zero width effects. Most exciting of
course would be the direct formation of sneutrinos
\cite{CC,godbole,DreLo}
\begin{equation}
e^+e^-\ra \ti{\nu}_\tau
\end{equation}
for sneutrino masses in the LEP2 range. The sneutrinos would manifest
themselves as a sharp resonance peak.
\section{Slepton Exchange in \ee Collisions}
At energies much lower than the sparticle masses, $R$-parity breaking
interactions introduce effective $llll$ and $llqq$ contact
interactions.  These operators will in general mediate $L$ violating
processes and FCNC processes so that existing data put stringent
constraints on the couplings.  However, if only some of the operators
with a particular generation structure are present in Eq.\ 
(\ref{superp}), then the effective four-fermion Lagrangian does not
violate lepton number.  Similarly, the couplings can be arranged such
that there are no other sources of FCNC interactions than CKM mixing
in the quark sector. In the purely leptonic sector, we can restrict
ourselves to the following two possibilities\footnote{There is only
  one additional possibility for which the effective four-lepton
  Lagrangian does not violate lepton flavor and that involves three
  Yukawa couplings, each violating all three lepton flavors. However,
  this case is less interesting experimentally as shown later.}:
\\[1mm]
(a) one single Yukawa coupling is much larger than all the others, so
that the latter can be neglected; \\[1mm]
(b) two Yukawa couplings are much larger than all the others, where
both couplings violate {\it one and the same} lepton flavor, or both
couplings violate {\it all three} lepton flavors.\\[2mm]
In these cases low-energy experiments are not restrictive and
typically allow for couplings $\lambda \lsim 0.1\times$($\ti{m}$/200
GeV), where $\ti{m}$ is the mass scale of the sparticles participating
in the process.  The corresponding limits, derived by assuming only
one non-vanishing coupling at a time, are summarized in
Table~\ref{tablam}.  The most stringent limits for $\lambda$ can be
derived from CC universality, lepton universality and the induced $\nu_e$
Majorana mass \cite{DH,CC}. Additional constraints on products of
$\lambda$ and $\lambda'$ couplings come from rare $K$ and $B$ leptonic
decay processes \cite{roy,jang}. In Table \ref{tablam} we include
those limits which are relevant for the present study. Examining all
possible combinations of $\lambda$ and $\lambda'$ couplings compatible
with these bounds, it turns out that if the HERA data are interpreted
as {\it top} squark production (\ie $\lambda'_{131} \gsim 0.05$), then
the $\lambda$-couplings relevant for leptonic processes are very strongly
constrained, $\lambda_{121} < 0.0036$, $\lambda_{131} < 0.04$, and
$\lambda_{123} < 0.048$. As a result, the effects on purely leptonic
processes at LEP2 due to slepton exchanges would be small. However, if
the HERA events are due to {\it charm} squark production (\ie
$\lambda'_{121} \gsim 0.05$), the rare $B$ and $K$ decays do not impose
strong constraints on $\lambda_{131}$ or $\lambda_{123}$ and we may
expect large effects due to $\tau$-sneutrino exchanges at LEP2.

\setlength{\doublerulesep}{3mm}
\begin{table}[htbp]
\begin{center}
\begin{tabular}{|c|c|c|c|c|c|c|c|c|c|}\hline \rule{0mm}{5mm}
$\lambda$& 121& 122& 123& 131& 132& 133& 231& 232& 233\\[1mm]
\hline \rule{0mm}{5mm}
Limit& 0.08$^a$& 0.08$^a$& 0.08$^a$& 0.20$^b$& 0.20$^b$& 0.006$^c$&
       0.18$^d$& 0.18$^d$& 0.18$^d$ \\
\hline\hline
\multicolumn{3}{|c|}{\rule{0mm}{7mm} Deacy mode}& 
\multicolumn{4}{c|}{Combinations constrained} &
\multicolumn{3}{c|}{Limit}\\[1mm]
\hline 
\multicolumn{3}{|c|}{\rule{0mm}{5mm} $K\ra e^{\pm}\mu^{\mp}$}
       & \multicolumn{4}{c|}{$\lambda_{121}\lambda'_{121}$}
       &\multicolumn{3}{c|}{$ 10^{-7}$}\\[1mm]
\multicolumn{3}{|c|}{\rule{0mm}{5mm} $B_d\ra e^{\pm}\mu^{\mp}$}
       & \multicolumn{4}{c|}{$\lambda_{121}\lambda'_{131}$}
       &\multicolumn{3}{c|}{$1.8\times 10^{-4}$}\\[1mm]
\multicolumn{3}{|c|}{\rule{0mm}{5mm}$B_d\ra e^{\pm}\tau^{\mp}$}
       & \multicolumn{4}{c|}{$\lambda_{131}\lambda'_{131}$}
       &\multicolumn{3}{c|}{$2.0\times 10^{-3}$}\\[1mm]
\multicolumn{3}{|c|}{\rule{0mm}{5mm}$B_d \ra \mu^{\pm}\tau^{\mp}$}
       & \multicolumn{4}{c|}{$\lambda_{123}\lambda'_{131}$}
       &\multicolumn{3}{c|}{$2.4\times 10^{-3}$}\\[1mm]
\hline
\end{tabular}
\caption{\it Upper part: 
  The 1$\sigma$ limits on the $R$-parity breaking couplings $\lambda$
  [in units of $\ti{m}$/200 GeV, where $\ti{m}$ is the appropriate
  sfermion mass], from (a) charged-current universality; (b)
  $\Gamma(\tau\ra e\nu\bar{\nu})/ \Gamma(\tau\ra\mu\nu\bar{\nu})$; (c)
  the induced $\nu_e$ Majorana mass; (d) $\Gamma(\tau\ra
  e\nu\bar{\nu})/ \Gamma(\mu\ra e\nu\bar{\nu})$; (a), (b) and (d) from
  Ref.~\protect\cite{CC}, (c) from Ref.~\protect\cite{DH}. Lower
  part: Limits on the products of $\lambda$ and $\lambda'$ which are
  relevant for our discussion [in units of ($\ti{m}$/200 GeV)$^2$];
  $K$ decay limits from Ref.~\protect\cite{roy}, $B$ decay limits
  from Ref.~\protect\cite{jang}.}
\label{tablam}
\end{center} 
\end{table}
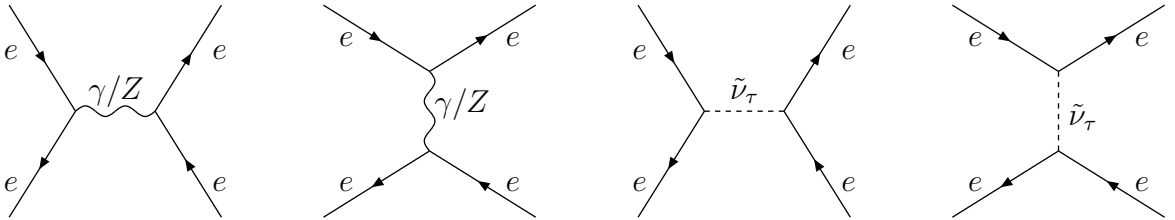
\begin{figure}[htbp]
%
\begin{picture}(100,100)(0,0)
\ArrowLine(10,90)(35,50)
\ArrowLine(35,50)(10,10)
\Photon(35,50)(65,50){2}{2}
\ArrowLine(90,10)(65,50)
\ArrowLine(65,50)(90,90)
\put(8,20){$e$}
\put(8,70){$e$}
\put(40,55){$\gamma/Z$}
\put(87,20){$e$}
\put(87,70){$e$}
\end{picture}
%
\begin{picture}(100,100)(-15,0)
\ArrowLine(10,90)(50,65)
\ArrowLine(50,65)(90,90)
\Photon(50,65)(50,35){2}{2}
\ArrowLine(90,10)(50,35)
\ArrowLine(50,35)(10,10)
\put(16,20){$e$}
\put(16,75){$e$}
\put(52,50){$\gamma/Z$}
\put(79,20){$e$}
\put(79,75){$e$}
\end{picture}
%
\begin{picture}(100,100)(-30,0)
\ArrowLine(10,90)(35,50)
\ArrowLine(35,50)(10,10)
\DashLine(35,50)(65,50){2}
\ArrowLine(90,10)(65,50)
\ArrowLine(65,50)(90,90)
\put(8,20){$e$}
\put(8,70){$e$}
\put(45,55){$\tilde{\nu}_{\tau}$}
\put(87,20){$e$}
\put(87,70){$e$}
\end{picture}
%
\begin{picture}(100,100)(-45,0)
\ArrowLine(10,90)(50,65)
\ArrowLine(50,65)(90,90)
\DashLine(50,65)(50,35){2}
\ArrowLine(90,10)(50,35)
\ArrowLine(50,35)(10,10)
\put(16,20){$e$}
\put(16,75){$e$}
\put(54,45){$\tilde{\nu}_{\tau}$}
\put(79,20){$e$}
\put(79,75){$e$}
\end{picture}
\caption{\label{baba} \it Diagrams for Bhabha scattering
  $e^+e^- \rightarrow e^+ e^-$ including $s$- and $t$-channel exchange
  of $\tilde{\nu}_{\tau}$ ($\lambda_{131} \ne 0$).}  
\end{figure}
\begin{figure}[htbp]
%
\begin{picture}(100,100)(-100,0)
\ArrowLine(10,90)(35,50)
\ArrowLine(35,50)(10,10)
\Photon(35,50)(65,50){2}{2}
\ArrowLine(90,10)(65,50)
\ArrowLine(65,50)(90,90)
\put(8,20){$e$}
\put(8,70){$e$}
\put(40,55){$\gamma/Z$}
\put(87,20){$\tau$}
\put(87,70){$\tau$}
\end{picture}
%
\begin{picture}(100,100)(-150,0)
\ArrowLine(10,90)(50,65)
\ArrowLine(50,65)(90,90)
\DashLine(50,65)(50,35){2}
\ArrowLine(90,10)(50,35)
\ArrowLine(50,35)(10,10)
\put(16,20){$e$}
\put(16,75){$e$}
\put(54,45){$\tilde{\nu}_e$}
\put(79,20){$\tau$}
\put(79,75){$\tau$}
\end{picture}
\caption{\label{mumu} \it Diagrams for $e^+e^- \rightarrow
  \tau^+ \tau^-$ including $t$-channel exchange of $\tilde{\nu}_e$
  ($\lambda_{131} \ne 0$).}
\end{figure}
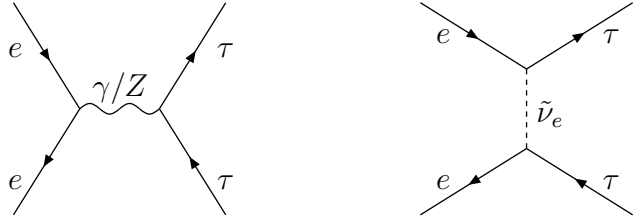

We will first consider the {\it case} (a) taking specifically
$\lambda_{131} \ne 0$.  The cross section for Bhabha scattering is
then built up by the $s$- and $t$-channel exchange of $\gamma,Z$
bosons and of (anti-)sneutrinos $\ti{\nu}_\tau$,
$\bar{\tilde{\nu}}_{\tau}$ (see Fig.~\ref{baba}). The cross section
can be written most transparently in terms of helicity amplitudes
\cite{SZ}:
\begin{eqnarray}
&&\frac{\mbox{d}\sigma}{\mbox{d}\cos\theta} (e^+ e^- \ra
 e^+e^-) = \non \\[2ex]
&&\frac{\pi\alpha^2s}{8}
\Bigl\{(1+\cos\theta)^2 
\left[ |f^s_{LR}|^2 + |f^s_{RL}|^2 +
  |f^t_{LR}|^2 + |f^t_{RL}|^2 - 2\mbox{Re}(f^{s\;*}_{LR}\,f^t_{LR}) -
  2\mbox{Re}(f^{s\;*}_{RL}\,f^t_{RL}) \right] \non \\[1.5ex]
&&+ (1-\cos\theta)^2\left[|f^s_{LL}|^2+|f^s_{RR}|^2\right] +
  4\left[|f^t_{LL}|^2+|f^t_{RR}|^2\right] \Bigr\}
\label{sigbb}
\end{eqnarray}
While the $s$- and $t$-channel $\gamma,Z$ amplitudes in the Standard
Model involve the coupling of vector currents, the sneutrino exchange
is described by scalar currents. By performing appropriate Fierz
transformations, the $s$-channel $\ti{\nu}$ exchange amplitudes can 
be rewritten, however, as $t$-channel vector amplitudes, and
$t$-channel $\ti{\nu}$ exchange amplitudes as $s$-channel vector
amplitudes:
\beq 
(\bar{e}_R e_L)(\bar{e}'_Le'_R) \ra
-\frac{1}{2}(\bar{e}_R\gamma_{\mu} e'_R)(\bar{e}'_L\gamma_{\mu}e_L)
\eeq 
The independent $s$-channel amplitudes $f^s_{h_i h_f}$
are therefore given by 
\beq 
f^s_{LR} &=& \frac{1}{s} +
\frac{g_L^2}{s-m^2_Z+i\Gamma_Zm_Z}
\label{fslr}\\ 
f^s_{RL} &=& \frac{1}{s} + \frac{g_R^2}{s-m^2_Z+i\Gamma_Zm_Z}
\label{fsrl}\\ 
f^s_{LL} &=& \frac{1}{s} + \frac{g_Lg_R}{s-m^2_Z+i\Gamma_Zm_Z}
+\frac{1}{2}\frac{(\lambda_{1j1}/e)^2}{t-m^2_j}
\label{fsll}\\ 
f^s_{RR} &=&
\frac{1}{s} +\frac{g_Lg_R}{s-m^2_Z+i\Gamma_Zm_Z}
+\frac{1}{2}\frac{(\lambda_{1j1}/e)^2}{t-m^2_j} 
\label{fsrr}
\eeq
\begin{figure}[htbp] 
  \unitlength 1mm
\begin{picture}(162,160)
  \put(15,19){ \epsfxsize=32.3cm \epsfysize=34cm \epsfbox{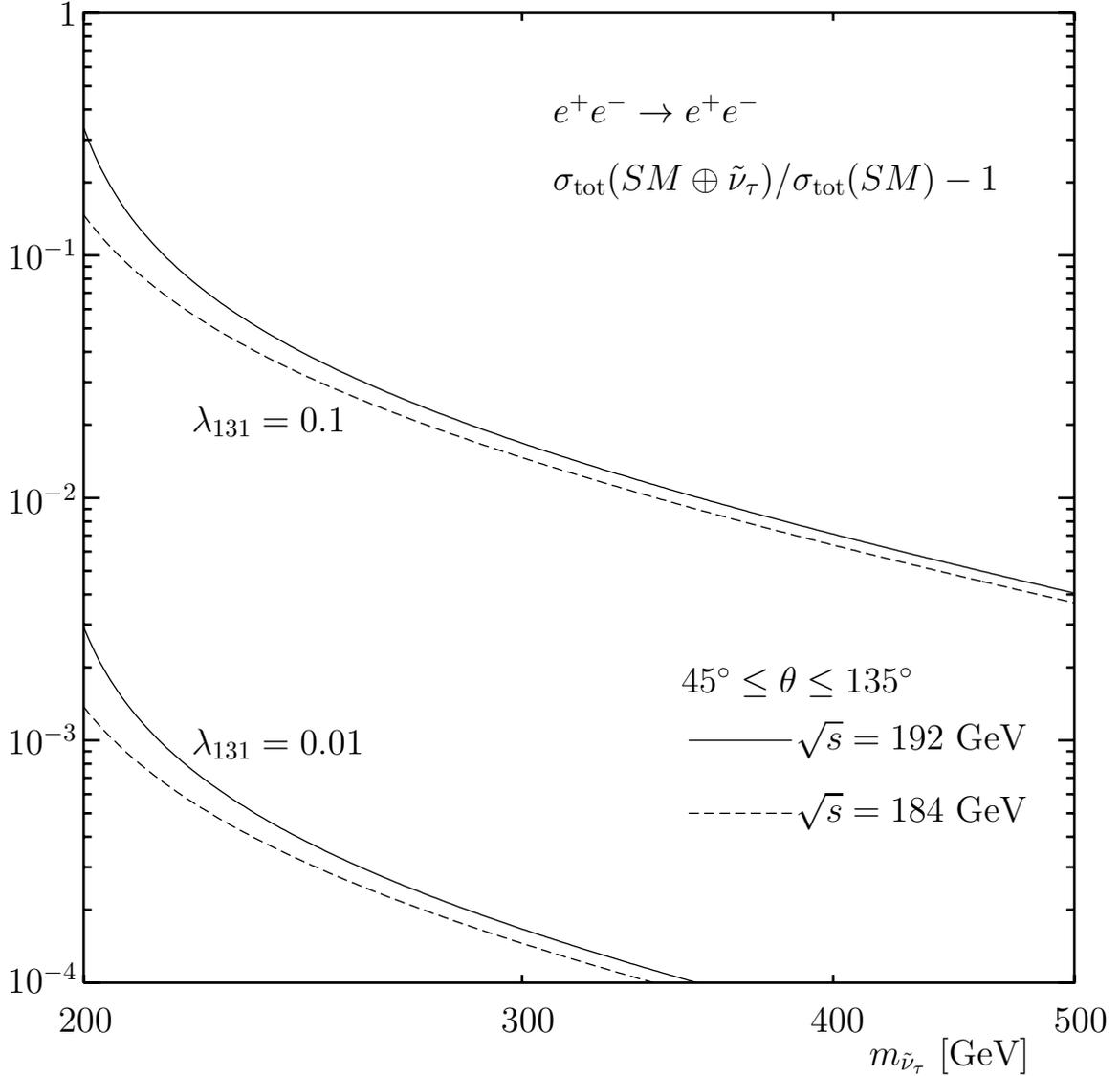}}
  \put(114.6,34){\makebox(0,0)[l]{\large $\protect\sqrt{s} = 184$
      GeV}} 
\put(114.6,44){\makebox(0,0)[l]{\large $\protect\sqrt{s} = 192$ GeV}} 
\put(98.7,52){\makebox(0,0)[l]{\large $45^{\circ} \leq \theta \leq
    135^{\circ}$}} 
\put(30.7,43.4){\makebox(0,0)[l]{\large $\lambda_{131} = 0.01$}} 
\put(30.7,88.2){\makebox(0,0)[l]{\large $\lambda_{131} = 0.1$}} 
\put(80.9,121.7){\makebox(0,0)[l]{\large $\sigma_{\rm tot}(SM\oplus
    \tilde{\nu}_{\tau})/\sigma_{\rm  tot}(SM)-1$}} 
\put(80.9,131.7){\makebox(0,0)[l]{\large $e^+e^- \rightarrow e^+e^-$}} 
\put(136,0){\makebox(0,0){\large $m_{\tilde{\nu}_{\tau}}~$[GeV]}} 
\put(14.7,145){\makebox(0,0)[r]{\large 1}}
\put(14.7,111.5){\makebox(0,0)[r]{\large $10^{-1}$}}
\put(14.7,77.5){\makebox(0,0)[r]{\large $10^{-2}$}}
\put(14.7,44){\makebox(0,0)[r]{\large $10^{-3}$}}
\put(14.7,10.8){\makebox(0,0)[r]{\large $10^{-4}$}}
\put(12.2,5){\makebox(0,0)[l]{\large 200}}
\put(73.7,5){\makebox(0,0)[l]{\large 300}}
\put(116.7,5){\makebox(0,0)[l]{\large 400}}
\put(150.5,5){\makebox(0,0)[l]{\large 500}}
\end{picture}
\caption{\it 
  Effect of sneutrino $\ti{\nu}_{\tau}$ exchange on the cross section
  for Bhabha scattering for $45^{\circ} \leq \theta \leq 135^{\circ}$
  at $\protect\sqrt{s} = 192$ GeV (full lines) and $\protect\sqrt{s} =
  184$ GeV (dashed lines).}
\label{figbb}
\end{figure}
In the same way, the $t$-channel exchange amplitudes $f^t_{h_i h_f}$ 
can be written as 
\begin{eqnarray}
f^t_{LR} &=& \frac{1}{t} + \frac{g_L^2}{t-m^2_Z}
\label{ftlr}\\ 
f^t_{RL} &=& \frac{1}{t} + \frac{g_R^2}{t-m^2_Z}
\label{ftrl}\\ 
f^t_{LL} &=& \frac{1}{t} + \frac{g_Lg_R}{t-m^2_Z}
+\frac{1}{2}\frac{(\lambda_{1j1}/e)^2}{s-m^2_j+i\Gamma_j m_j}
\label{ftll}\\ 
f^t_{RR} &=& \frac{1}{t} +\frac{g_Lg_R}{t-m^2_Z}
+\frac{1}{2}\frac{(\lambda_{1j1}/e)^2}{s-m^2_j+i\Gamma_j m_j}
\label{ftrr}
\end{eqnarray} 
The parameters $m_j$ and $\Gamma_j$ are the mass and width of the
sneutrino $\ti{\nu}_j = \tilde{\nu}_{\tau}$.  To simplify notations we
have defined the indices $L,R$ to denote the helicities of the {\it
  ingoing electron} (first index) and the {\it outgoing positron}
(second index).  The helicities of the ingoing positron and the
outgoing electron are fixed by the $\gamma_5$ invariance of the vector
interactions: they are opposite to the helicities of the lepton
partner in $s$-channel amplitudes and the same in $t$-channel
amplitudes\footnote{The ($LR$) and ($RL$) terms of the first line of
  Eq.\ (\ref{sigbb}) correspond to equal electron helicities in the
  initial and final state so that forward scattering is permitted;
  this is obvious for the $s$-channel amplitudes, but applies also to
  the $t$-channel amplitudes after an appropriate Fierz
  transformation. The first two terms of the second line correspond to
  opposite electron helicities so that forward scattering is
  forbidden. Finally, the last two terms correspond to isotropic
  spin-zero scattering which becomes apparent after applying a Fierz
  transformation from the $t$- to the $s$-channel.}. The left/right
$Z$ charges\footnote{Note that in Eqs.~(\ref{fslr}-\ref{ftrr}) the
  outgoing positron with the helicity $L(R)$ couples with the charge
  $g_R (g_L)$.} of the leptons are defined as
\begin{eqnarray}
g_L&=&\left(\frac{\sqrt{2}G_{\mu}m^2_Z}{\pi\alpha}\right)^{1/2}
\left[I_3^l-s^2_W Q^l\right] \non\\ 
g_R&=&\left(\frac{\sqrt{2}G_{\mu}m^2_Z}{\pi\alpha}\right)^{1/2}
\left[  {} -s^2_W Q^l\right]\non 
\end{eqnarray}

In Fig.~\ref{figbb} the impact of the sneutrino $\ti{\nu}_{\tau}$
exchange on the Bhabha scattering process at LEP2 energies  is shown
as a function of the sneutrino mass, assuming couplings
$\lambda_{131}=0.1$ or $\lambda_{131} = 0.01$.  Due to the $s$-channel
exchange, the effect can be very large if the sneutrino mass is
close to the LEP2 center-of-mass energy.

\begin{figure}[htbp] 
  \unitlength 1mm
\begin{picture}(162,160)
\put(15,19){ 
\epsfxsize=32.3cm 
\epsfysize=34cm 
\epsfbox{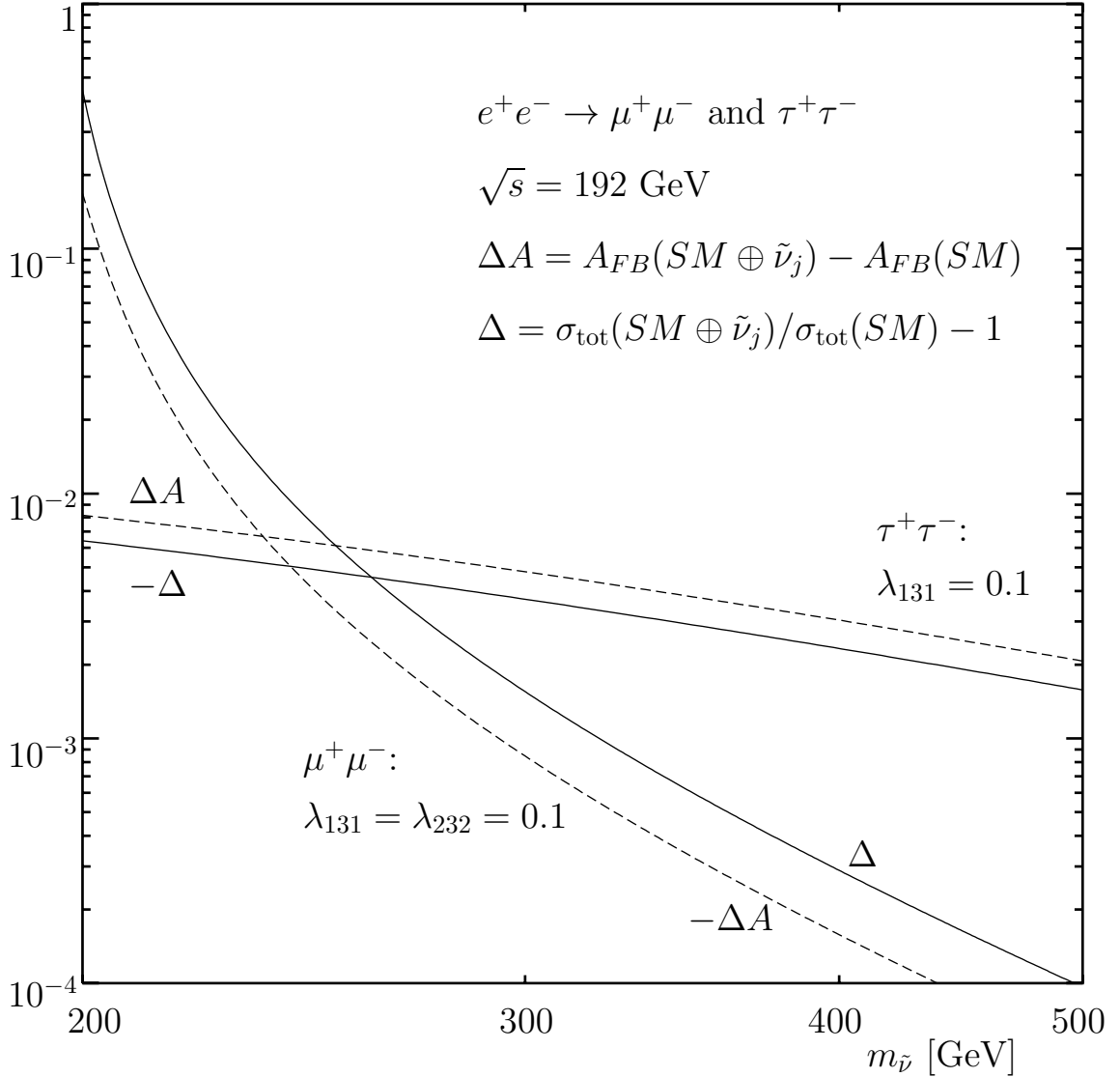}}
\put(70,110){\makebox(0,0)[l]{\large $\Delta A= A_{FB}(SM
    \protect\oplus\tilde{\nu}_j) - A_{FB}(SM)$}} 
\put(70,100){\makebox(0,0)[l]{\large $\Delta = \sigma_{\rm tot}(SM
    \protect\oplus\tilde{\nu}_j) / \sigma_{\rm tot}(SM) - 1$}} 
\put(46,41){\makebox(0,0)[l]{\large $\mu^+\mu^-$:}}
\put(46,33){\makebox(0,0)[l]{\large $\lambda_{131}=\lambda_{232} = 0.1$}}
\put(125,73){\makebox(0,0)[l]{\large $\tau^+\tau^-$:}}
\put(125,65){\makebox(0,0)[l]{\large $\lambda_{131} = 0.1$}}
\put(121,28){\makebox(0,0)[l]{\large $\Delta$}}
\put(22,65){\makebox(0,0)[l]{\large $-\Delta$}}
\put(22,78){\makebox(0,0)[l]{\large $\Delta A$}}
\put(99,19){\makebox(0,0)[l]{\large $-\Delta A$}}
\put(70,120){\makebox(0,0)[l]{\large $\protect\sqrt{s} = 192$ GeV}}
\put(70,130){\makebox(0,0)[l]{\large $e^+e^- \rightarrow \mu^+\mu^-$
    and $\tau^+\tau^-$}}
\put(134,0){\makebox(0,0){\large $m_{\tilde{\nu}}$~[GeV]}}
\put(15,143){\makebox(0,0)[r]{\large 1}}
\put(15,109.5){\makebox(0,0)[r]{\large $10^{-1}$}}
\put(15,76){\makebox(0,0)[r]{\large $10^{-2}$}}
\put(15,42){\makebox(0,0)[r]{\large $10^{-3}$}}
\put(15,9){\makebox(0,0)[r]{\large $10^{-4}$}}
\put(21,5){\makebox(0,0)[r]{\large 200}}
\put(80.5,5){\makebox(0,0)[r]{\large 300}}
\put(123,5){\makebox(0,0)[r]{\large 400}}
\put(156.5,5){\makebox(0,0)[r]{\large 500}}
\end{picture}
\caption{\it Effect of sneutrino exchange on $e^+e^- \rightarrow
  \ell^+\ell^-$ in two different scenarios: the curves labeled by
  $\lambda_{131} = 0.1$ correspond to a scenario with the additional
  $t$-channel exchange of $\tilde{\nu}_e$; the curves labeled by
  $\lambda_{131} = \lambda_{232} = 0.1$, with additional $s$-channel
  exchange of $\ti{\nu}_{\tau}$.  Full lines: $\Delta =
  \sigma(SM\protect\oplus \tilde{\nu}_j) / \sigma(SM) - 1$, dashed
  lines $\Delta A = A_{FB}(SM \protect\oplus \tilde{\nu}_j) -
  A_{FB}(SM)$ for $\protect\sqrt{s} = 192$ GeV .}
\label{figmm}
\end{figure}

The analysis of $\tau^+\tau^-$ production in \ee annihilation proceeds
in an analogous way. An important difference is the absence of the
$t$-channel Standard Model amplitude and the $s$-channel sneutrino
exchange amplitude if only the Yukawa coupling $\lambda_{131}$ is
assumed to be non-zero (\ie $\lambda_{1j1} = \lambda_{131}$ in Eqs.\ 
(\ref{fsll}, \ref{fsrr}) and $f^t = 0$ in Eqs.\ 
(\ref{ftlr}-\ref{ftrr})).  In this case the $\tau^+\tau^-$ production
process is mediated by the $s$-channel $\gamma,Z$ exchange and the
exchange of the (anti-)sneutrino $\ti{\nu}_e$ in the $t$-channel (see
Fig.~\ref{mumu}).  Because the $s$-channel sneutrino exchange diagram
is absent, the impact on the total cross section is small even for
$\lambda_{131}$ as large as 0.1 as can be seen in Fig.~\ref{figmm}. In
the scenario considered, that is for $\lambda_{131} \ne 0$ and all
other Yukawa couplings vanishing, the process $e^+e^- \ra \mu^+\mu^-$
is not affected and the cross section is given by the Standard Model.

Given the bounds of Table \ref{tablam}, electron sneutrino exchange
involving $\lambda_{121}$ cannot contribute to $\mu$ pair production
in this specific $\lambda$ scenario.

Finally, in the realization of case (a) with $\lambda_{123} \ne 0$ all
lepton flavors are violated. Bhabha scattering is then not affected at
all. However, \mm\ and \taus\ pair production in \ee scattering would
receive contributions from $t$-channel $\tilde{\nu}_{\tau}$ and
$\tilde{\nu}_{\mu}$ sneutrino exchanges, respectively (\ie 
$\lambda_{1j1} \Rightarrow \lambda_{123}$ in Eqs.\ (\ref{fsll},
\ref{fsrr}) and $f^t = 0$ in Eqs.\ (\ref{ftlr}-\ref{ftrr})).

{\it Case} (b) with two large Yukawa couplings is interesting if both
couplings violate the same lepton flavor: $\lambda_{131}$
and $\lambda_{232}\ne 0$, for example. If this scenario is realized, then the
process $e^+e^-\ra \mu^+\mu^-$ receives an additional contribution
from $s$-channel $\ti{\nu}_{\tau}$ sneutrino exchange.  Therefore,
$\mu^+\mu^-$ production would be affected in a similar way as Bhabha
scattering, which is apparent from  Fig.~\ref{figmm}.  

Stringent bounds on contact interactions in the lepton sector have
been reported by the LEP experiments \cite{Richard,13a}.  Defining the
contact interactions by the Lagrangian
\begin{equation}
{\cal L}_{CI}^{f,ij} = 
\pm \frac{4\pi}{\Lambda_{ij}^2} 
(\bar{e}_i \gamma_{\mu} e_i) (\bar{f}_j \gamma_{\mu}f_j)
\end{equation}
with $i,j = L,R$, the lower bounds for the $LR$ and $RL$ scales and
the positive sign are close to 2.7 TeV, while for the negative sign
they are close to 3.2 TeV ($LL$ and $RR$ bounds are even stronger).
Even though these values cannot be transferred immediately to the more
complex analysis presented here, we nevertheless expect typical values
of $m_{\ti{\nu}}/\lambda\sim \Lambda/\sqrt{8\pi}\simeq 0.5$ to 0.7 TeV
as an order of magnitude estimate in the present scenario.  Choosing
$m_{\ti{\nu}}\simeq 200$ GeV, the Yukawa couplings could still be of
the order 0.4. This analysis is based on an integrated luminosity of
$\int {\cal L} \sim 10$ pb${}^{-1}$ at $\sqrt{s} = 161$ GeV. Since the
limits on $\Lambda$ scale with $\left(\int{\cal L}\right)^{1/4}$
\cite{Treille}, improvements by a factor $\sim 2.5$ can be expected
for a total integrated luminosity of $\int{\cal L}=400$ pb${}^{-1}$,
which can be anticipated for the 4 combined LEP experiments in the
runs of this year. The sensitivity on the scale of the contact
interactions in the lepton sector will then rise to a value close to
$\Lambda \sim 7$ to 8 TeV corresponding to $\lambda \simeq 0.13$ to
0.15 $\times (m_{\tilde{\nu}}/200$ GeV).

\section{Resonance Formation}
The most exciting prediction of $R$-breaking supersymmetry in the
lepton sector, however, is the formation of sneutrino resonances
\cite{CC,godbole,DreLo} with masses either close to the LEP2 energy or
even in reach of the machine. The production of $\tilde{\nu}_{\tau}$
sneutrinos would be compatible with all low-energy constraints known
so far, $e^+e^-\ra \ti{\nu}_{\tau}$.  If the HERA high $x$, high $Q^2$
data indeed indicate the production of a 2nd generation squark
$\ti{c}$, sneutrinos may also exist in the mass range around 200 GeV.
Na{\"\i}vely one would expect non-colored states to be lighter than
the associated colored states.  Even if the stop $\ti{t}_1$ mass is
reduced through strong left-right mixing by the large Yukawa
interactions in the $t,\ti{t}$ sector, in a large part of the
supersymmetric parameter space the sneutrino masses can be as light as
200 GeV in grand unified models incorporating universal soft SUSY
breaking parameters (see Ref.\ \cite{masses} for example).

\begin{figure}[htbp] 
  \unitlength 1mm
\begin{picture}(162,160)
  \put(15,19){ \epsfxsize=32.3cm \epsfysize=34cm
\epsfbox{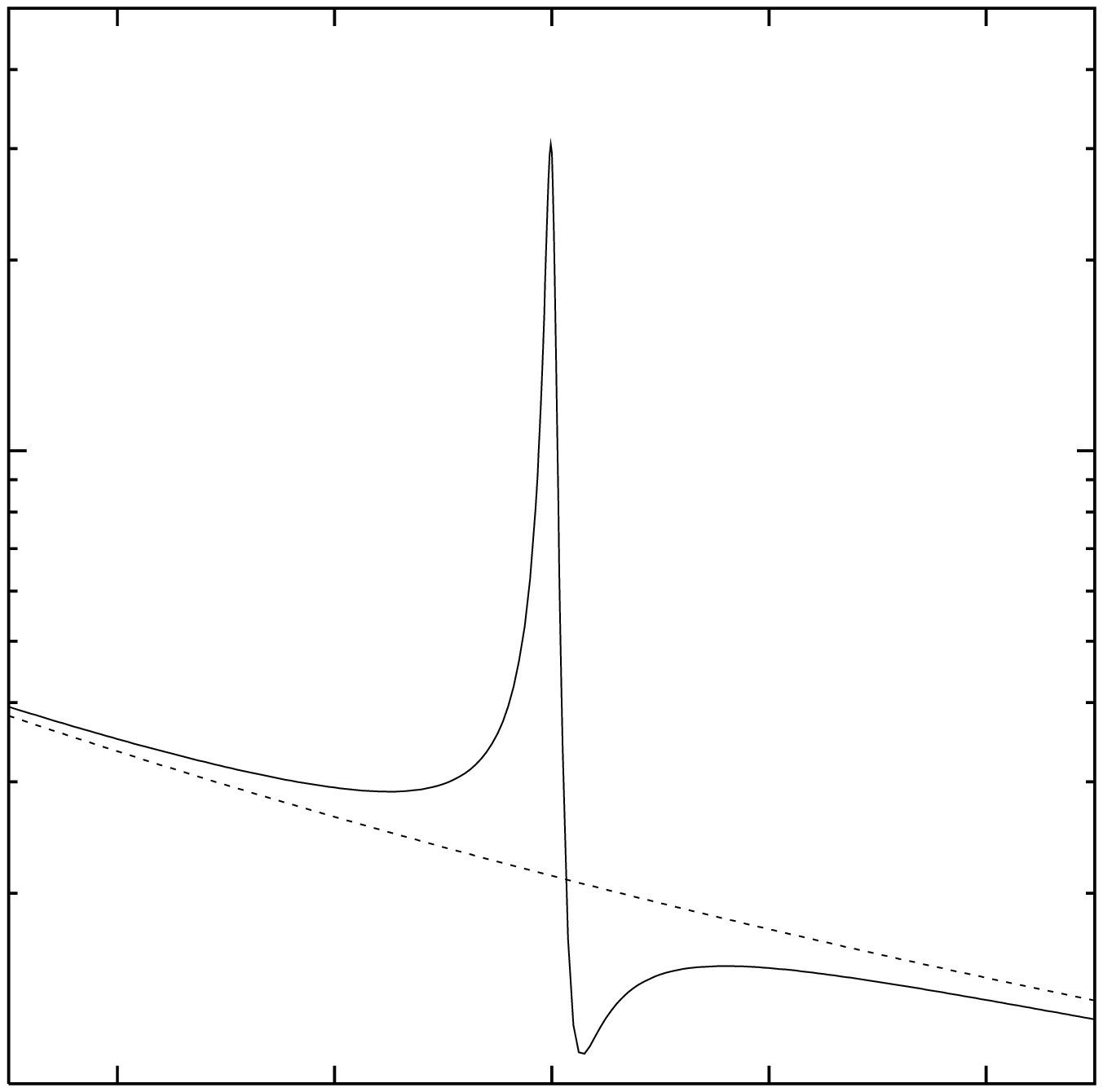}}
\put(31.2,100){\makebox(0,0)[l]{\large $m_{\tilde{\nu}} = 200$ GeV}}
\put(31.2,92){\makebox(0,0)[l]{\large $\Gamma_{\tilde{\nu}} = 1$ GeV}}
\put(31.2,84){\makebox(0,0)[l]{\large $\lambda_{131} = 0.1$}}
\put(96,121.5){\makebox(0,0)[l]{\large $45^{\circ} \leq \theta \leq
    135^{\circ}$}}
\put(96,128.4){\makebox(0,0)[l]{\large $\sigma_{\rm tot}(e^+e^-
    \rightarrow e^+e^-)$ [pb]}} 
\put(126.3,0){\makebox(0,0){\large $\protect\sqrt{s}$[GeV]}} 
\put(139.3,5.3){\makebox(0,0){\large 240}}
\put(112.0,5.3){\makebox(0,0){\large 220}} 
\put(84.6,5.3){\makebox(0,0){\large 200}}
\put(57.1,5.3){\makebox(0,0){\large 180}} 
\put(29.5,5.3){\makebox(0,0){\large 160}}
\put(14.0,89.6){\makebox(0,0)[r]{\large $10^2$}}
\put(14.0,10.0){\makebox(0,0)[r]{\large $10$}}
\end{picture}
\caption{\it 
  Cross section for Bhabha scattering including $\tilde{\nu}_{\tau},
  \bar{\tilde{\nu}}_{\tau}$ sneutrino resonance
  formation for $45^{\circ} \leq \theta \leq 135^{\circ}$ as a
  function of the $e^+e^-$ center-of-mass energy. Parameters:
  $m_{\tilde{\nu}} = 200$ GeV, $\Gamma_{\tilde{\nu}} = 1$ GeV, and
  $\lambda_{131} = 0.1$.}
\label{figres}
\end{figure}

The cross section for the production of sneutrinos which decay to a
specified final state $F$, is given by the Breit-Wigner formula
\begin{equation}
\sigma(e^+e^-\ra \ti{\nu}\ra F) = \frac{4\pi s}{m^2_{\ti{\nu}}}\;
\frac {\Gamma(\ti{\nu}\ra e^+e^-)\Gamma(\ti{\nu}\ra F)}
{(s-m^2_{\ti{\nu}})^2+m_{\ti{\nu}}^2\Gamma^2_{\ti{\nu}}} 
\end{equation}
The partial decay width $\Gamma(\ti{\nu}\ra e^+e^-)=\lambda^2_{1j1}
m_{\ti{\nu}} /16\pi$ is very small. However sneutrinos can also decay
via $R$-parity conserving gauge couplings to $\nu\chi^0$ and
$l^{\pm}\chi^{\mp}$ pairs with subsequent $\chi^0$ and $\chi^{\pm}$
decays and via $R$-parity violating $\lambda'$ couplings to $q\bar{q}$
pairs.  The partial decay widths for these channels depend on the
specific choice of the supersymmetry breaking parameters.  In large
regions of the supersymmetry parameter space, the total decay width of
sneutrinos can be as large as 1 GeV, \ie\ significantly larger than
the energy spread\footnote{Assuming that the energy spread $\delta E$
  scales with the square of the total energy, $\delta E\sim 204$ MeV
  is expected at $\sqrt{s}=192$ GeV at LEP2 \cite{Wells}.} at LEP2.
In this case the interference with the background Standard Model
process must be taken into account if $F = e^+e^-$ or \taus. The cross
sections including these interference effects have been presented in
Eqs.\ (\ref{sigbb}) and (\ref{fslr}) to (\ref{ftrr}). A representative
example for the cross section of the process $e^+e^- \rightarrow
e^+e^-$ including $\tilde{\nu}_{\tau}$ resonance formation is
displayed in Fig.\ \ref{figres}. Since the width is wider than the
beam energy spread, the maximum of the cross section is given by the
unitarity limit $\sigma_{\rm max} = (8\pi/m^2_{\ti{\nu}}) B_e^2$ for
sneutrino and anti-sneutrino production added up. The cross section in
the peak region is therefore very large.  In addition to the $\ell^+
\ell^-$ final states one should expect many other final states
generated in $R$-parity conserving $\ti{\nu}$ decays. Examples are
'Zen events'
\begin{equation}
e^+e^- \rightarrow \tilde{\nu} \rightarrow \nu\tilde{\chi}_1^0
~~~~~{\rm etc.}
\end{equation}
with $R$-parity breaking $\tilde{\chi}_1^0$ decays, or isolated lepton
events 
\begin{equation}
e^+e^- \rightarrow \tilde{\nu} \rightarrow \ell \tilde{W}
\left[ \rightarrow W\tilde{\chi}_1^0\right]~~~~~{\rm etc.}
\end{equation}
in cascade decays. In addition one can also expect $R$-parity
violating decays to quark jets $\tilde{\nu} \rightarrow jj$
\cite{erler}. 

\section{Summary}
In this paper we have shown that if $R$-parity is broken by explicit
lepton number violating operators in the leptonic sector, distinctive
signals in $e^+e^-\ra e^+e^-$, \mm\ and \taus\ processes are
predicted. Motivated by a plausible explanation of the HERA events
involving the $R$-parity breaking $LQ\bar{D}$ operator, we have
analysed the impact of the $LL\bar{E}$ operator on these leptonic
processes. Interpreting the HERA data as {\it charm} squark
production, the operator that violates $\tau$-flavor is the most
interesting scenario for LEP2 physics.  If sleptons do exist in the
mass range of 200 GeV, the effect of the sneutrino exchanges at LEP2
could be very large. If sneutrino masses were within the reach of
LEP2, sneutrinos would manifest themselves through resonance formation
in $e^+e^-$ collisions.\\[3ex]

\noindent {\large \bf Acknowledgements}\\[2ex]
Numerical cross checks have been performed with the help of CompHEP
\cite{comphep} adapted to the $R$-parity violating supersymmetry.  We
are grateful to G.~Ross and J.S.~Lee for a discussion on bounds of the
couplings $\lambda$, and to P.~Wells for information on the LEP2 beam
energy spread.  Thanks for discussions go also to P. M\"attig and W.D.
Schlatter.  Communications by J.~Holt, J.~Kirkby, F.~Richard,
D.~Treille and D.~Zerwas are gratefully acknowledged.


\end{document}